\documentclass[aps,prl, twocolumn,secnumarabic,amssymb, figure, superscriptaddress,floatfix, tightenlines]{revtex4-1}
\usepackage{graphicx}

\usepackage{braket}
\usepackage{multirow}
\usepackage{gensymb}
\usepackage{booktabs}

\begin{document}

\title{La-Doped BaSnO$_3$ Electron Transport Layer for Perovskite Solar Cells}%
\author{Chang Woo Myung}
\affiliation{Department of Chemistry, Ulsan National Institute of Science and Technology (UNIST), Ulsan 44919, Korea}

\author{Geunsik Lee}
\affiliation{Department of Chemistry, Ulsan National Institute of Science and Technology (UNIST), Ulsan 44919, Korea}

\author{Kwang S. Kim}
\affiliation{Department of Chemistry, Ulsan National Institute of Science and Technology (UNIST), Ulsan 44919, Korea}

\email{kimks@unist.ac.kr}
\date{\today}%


\begin{abstract}
Due to the photo-instability and hysteresis of TiO$_2$ electron transport layer (ETL) in perovskite solar cells (PSCs), novel electron transport materials are highly demanded. Here, we show ideal band alignment between La-doped BaSnO$_3$ (LBSO) and methyl ammonium (MA) lead iodide perovskite (MAPbI$_3$). The CH$_3$NH$_3$PbI$_3$/La$_x$Ba$_{(1-x)}$SnO$_3$ interface forms a stable all-perovskite heterostructure. The selective band alignment is manipulated with band gap renormalization by La-doping on the Ba site. LBSO shows high mobility, photo-stability, and structural stability, promising the next generation ETL materials.
\end{abstract}

\maketitle

Solar energy is a highly efficient and eco-friendly source for the future energy harvesting. In recent years, PSCs based on inorganic/organic hybrid halide ABX$_3$ (A = Cs$^+$, CH$_3$NH$_3^+$ (MA), CHN$_2$H$_4^+$ (FA); B = Pb$^{2+}$; X = Cl$^-$, Br$^-$ or I$^-$) have shown a rapid progress to achieve over 20 $\%$\cite{7,33} solar cell efficiency, which is considered to be most promising large-scale thin-film solar energy materials\cite{2}. PSC owns many interesting physical properties including giant dielectric screening\cite{3},  bottleneck of hot phonon relaxation process\cite{4},  free excitonic state\cite{5}, defect tolerance\cite{34,35}, and polaron state\cite{6} with an ideal band gap, mobility, and optical absorbance. 

	TiO$_2$ has been the most popular ETL material for PSC devices. Both anatase and rutile TiO$_2$ have achieved high power conversion efficiency (PCE)\cite{7,8}. However, the ultraviolet-induced photocatalysis of TiO$_2$ shows device degradation and gradual performance deterioration over a long exposure\cite{9}. Direct optical measurements have shown that electron barrier of $\sim$ 0.1 eV may exist at the interface of TiO$_2$/MAPbI3\cite{10}, and such a large electron barrier causes \textit{I-V} hysteresis possibly due to accumulation of iodine defects at TiO$_2$/MAPbI$_3$\cite{11,12,13}. To overcome these obstacles, a development for an efficient and stable next generation ETL materials is recently emerging\cite{14}. Up to date, the metal oxide materials such as SnO$_2$, ZnO, WO$_3$, ZnSnO$_4$ etc. are being spotlighted as the next generation ETL\cite{15}. In a recent experiment, a new synthetic method for LBSO below 500 $^{\circ}$C was developed; photostable 5 mol$\%$ LBSO as ETL achieved PCE over 22 $\%$ with 1000 hours illumination\cite{16}. 
	
	BaSnO$_3$ (BSO) is known as a wide indirect band gap ($E_g$ $>$ 3.1 eV) material\cite{17,18}  with high electrical mobility\cite{18} $\sim$300 cm$^2$$\cdot$V$\cdot$s$^{-1}$ with possible existence of polaron state. Recent measurement of BSO thin film\cite{19} has achieved the room temperature conductivity over ~10$^4$ S$\cdot$cm$^{-1}$. The band gap ($E_g$) of BSO is easily tunable via chemical substitution with Pb, Bi or La\cite{17}. As proved by the hard X-ray photoelectron spectroscopy experiment\cite{20}, this band gap tuning could be explained by the band gap renormalization due to an electrostatic interaction between the dopant cation (La$^{3+}$) and extra free electrons in the conduction band\cite{21}. Therefore, rather than tuning the band gap of PSCs by A-site cation or halides substitutions, one can directly tune the band gap of ETL using BSO to achieve the optimal band alignment. We show that this scenario is also plausible at the interface of LBSO slab evidenced by a large energy difference ~ 2-4 eV between the conduction band (Sn, O-s states) and La states (d states), which is consistent with the bulk LBSO case proven by X-ray absorption spectroscopy measurement\cite{20}.
	
	 In this work, using the density functional theory (DFT), we unveil that La doping on BSO is the key factor for the optimal band alignment at the LBSO/MAPbI$_3$ interface. We found that though the terminations of both MAPbI$_3$ and LBSO can affect the band alignment, the hybridization between two layers removes the dependency on the terminations. We also found that the calculated binding energies between LBSO and PSCs with various terminations are large enough to stabilize the interface.
	 
  We used Vienna Ab initio Simulation Package (VASP)\cite{28} for non-collinear DFT calculations using PBE0 functional plus D3 van der Waals correction\cite{29} with inclusion of the dipole correction and the spin-orbit coupling (SOC) by switching off any presumed symmetry. We have employed the electrostatic potential of the solid to define the common vacuum level and to relate the calculation of PBE+SOC+D3 to that of PBE0+SOC+D3 which includes the Hartree-Fock exchange. The reference electrostatic local potential only includes the Hartree potential so that it is independent of the exchange-correlation functionals we choose \cite{30,31} (see the Supplemental Material).

	We employed a symmetric slab for BSO and MAPbI$_3$ slabs. A supercell consists of 3 BSO layers (BaO-terminated and SnO$_2$-terminated) and 3 layers of cubic MAPbI$_3$ layers (MAI-terminated and PbI$_2$-terminated) with (001) orientation for both, where the lattice mismatch using BSO(001)-$(3\times3)$ and MAPbI$_3$(001)-$(2\times2)$ is as small as ~ 1.0 $\%$. A vacuum size of ~ 30 $\AA$ is included.

\begin{figure}
\centering
\includegraphics[width=8.6cm]{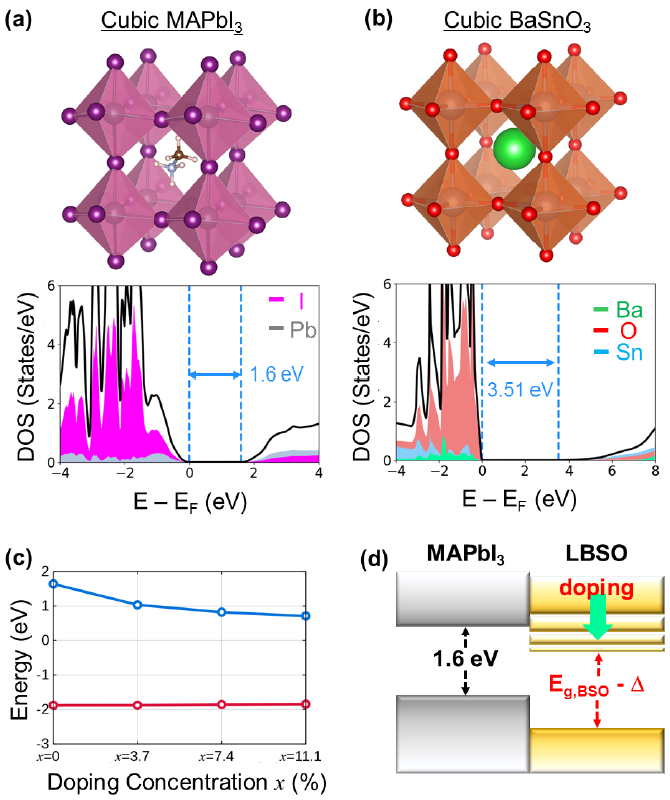}
\caption{\label{Figure. 1} Geometrical and electronic structures of MAPbI$_3$ and La-doped BaSnO$_3$, band gap renormalization upon La doping and band alignment diagram between MAPbI$_3$ and LBSO. (a) cubic MAPbI$_3$ perovskite crystal structure and PBE0+SOC density of states (DOS). (b) cubic BSO crystal structure and PBE0+SOC DOS. (c) The PBE0+SOC band gap change of LBSO with respect to La doping concentration $x$ = 0.0, 3.7, 7.4 and 11.1 $\%$. The local potential of each system has been set to E = 0 eV as a reference. (e) The schematic for the band alignment diagram between MAPbI$_3$ and La$_x$Ba$_{(1-x)}$SnO$_3$. Due to La doping, the band gap of BSO $E_g$ renormalizes to $E_g-\Delta$, where $\Delta > 0$.}
\end{figure}

We first investigate the electronic structures of bulk cubic MAPbI$_3$ (Fig. 1(a)) and cubic BaSnO$_3$ (Fig. 1(b)) perovskites using PBE0+SOC+D3 functional which correctly describes the experimental band gaps of MAPbI$_3$ (1.5–1.6 eV)\cite{22}  and BSO (3.1-3.5 eV)\cite{17}. Indeed, our calculation shows that PBE0+SOC+D3 is the optimal choice, E$_{g,PBE0+SOC+D3}$(MAPbI$_3$) = 1.59 eV (Fig. 1(a)) and E$_{g,PBE0+SOC+D3}$(BSO) = 3.51 eV (Fig. 1(b)). We find HSE06+SOC+D3 functional underestimates the experimental band gap for both materials, E$_{g,HSE06+SOC+D3}$(MAPbI$_3$) = 1.03 eV and E$_{g,HSE06+SOC+D3}$(BSO) = 2.79 eV. We also find that BSO's band gap is susceptible to the lattice constant $a$(BSO). Thus, we have used $a$(BSO) = 4.09 $\AA$ to match the experimental band gap. In contrast to MAPbI$_3$ in which the conduction and valence energy levels significantly shift, such shifts in BSO are almost negligible with the inclusion of SOC, $\Delta E_g$ = 40 meV. In BSO, the conduction band minimum (CBM) is the bonding state between Sn $s$-orbital and O $s$-orbitals at $\Gamma$ of Brillouin zone (BZ). VBM is the anti-bonding state of O $p$-orbitals at R of BZ (Fig. 1(b)). For MAPbI$_3$, The CBM states are $\ket{j=1/2}$ and $\ket{j=3/2}$ bonding states of Pb $p$-orbital and I $s$-orbital and the VBM state is $\ket{s=1/2}$ anti-bonding state of I $p$-orbital and Pb $s$-orbital (Fig. 1(a)). 

	The band level shift is only desirable in the conduction band, unless an elevated valence band level would allow hole carriers to transport and recombine with electron carriers in ETL materials. Indeed, we note that La-doping on BSO shifts the conduction band energy down, leaving the valence band energy intact (Fig. 1(c)). Upon La-doping ($x = 0, 3.7, 7.4, 11 \%$), the band gap of LBSO decreases such as $E_{g,x=0 \%}(3.51$ eV$) > E_{g,x=3.7\%}(2.91$ eV$) > E_{g,x=7.4\%}(2.68$ eV$) > E_{g,x=11.1\%}($2.55 eV) due to the electrostatic interaction between La$^{3+}$ dopant and the electrons in the conduction band (Fig. 1(d))\cite{17,18,20}.
	
\begin{figure}
\centering
\includegraphics[width=8.6cm]{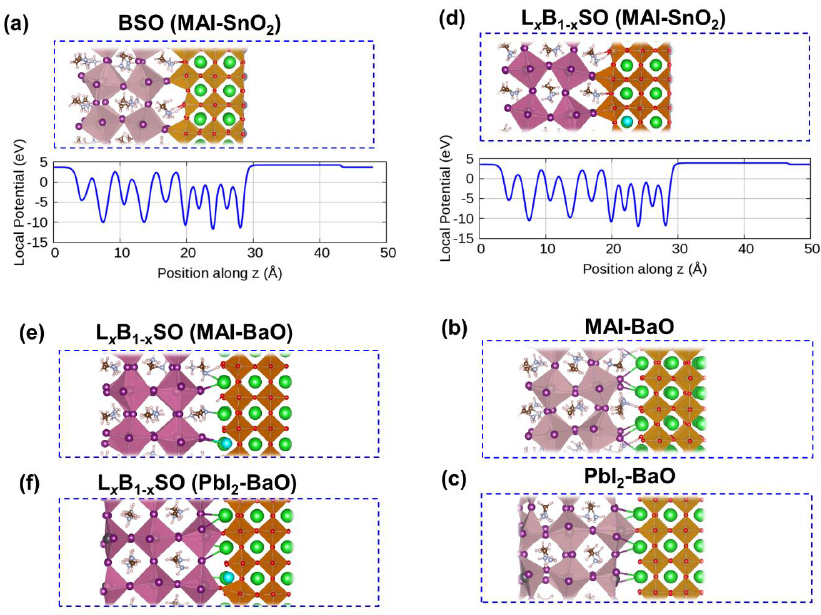}
\caption{\label{Figure. 2} DFT optimized BSO/MAPbI3 interface of (a) MAI-SnO$_2$, (b) MAI-BaO, (c) PbI$_2$-BaO termination and L$_x$B$_{(1-x)}$SO/MAPbI$_3$ ($x = 3.7 \%$) interface of (d) MAI-SnO$_2$, (e) MAI-BaO, (f) PbI$_2$-BaO termination. [Pb(gray), I(purple), C(brown), N(light blue), Sn(silver), O(red), Ba(green), and La(cyan)]. The integrated local electrostatic potential profile along z-direction of MAI-SnO2 interface is shown as an example. The kink appearing in the vacuum region indicates two different work functions of MAPbI$_3$ and L$_x$B$_{(1-x)}$SO by the dipole corrections.}
\end{figure}
	
	We studied various interfacial terminations between LBSO and MAPbI$_3$. Because the work function, band gap, and band alignment are significantly affected by the slab terminations, the study of various terminations are imperative for the thorough understanding of the interface. For MAPbI$_3$, experiments and DFT studies showed that MAI termination is thermodynamically more stable than PbI$_2$ termination\cite{23}. However, in PbI$_2$-rich condition PbI$_2$-terminated MAPbI$_3$ surface can also be formed\cite{24}. Previous DFT study revealed that the work function of MAI termination is $\sim$1 eV smaller than that of PbI$_2$ termination, indicating that the termination of MAPbI$_3$ slab plays a crucial role in determining the work function and its band alignment\cite{23}. Along (001), the interface between BSO and MAPbI$_3$ has four possible morphologies: MAI-terminated MAPbI$_3$ and SnO$_2$-terminated BSO (MAI-SnO$_2$) (Fig. 2(a)), MAI-terminated MAPbI$_3$ and SnO$_2$-terminated BSO (MAI-SnO$_2$) (Fig. 2(b)), PbI$_2$-terminated MAPbI$_3$ and BaO-terminated BSO (PbI$_2$-BaO) (Fig. 2(c)) and PbI$_2$-terminated MAPbI$_3$ and SnO$_2$-terminated BSO (PbI$_2$-SnO$_2$) (see the Supplemental Material). We investigated the minimum energy configurations of the interfaces and corresponding binding energies E$_b$. For the PbI$_2$-SnO$_2$ interface, the interfacial PbI$_2$ (in perovskite form) becomes unstable and forms PbO$_2$ with oxygen anions of SnO$_2$ termination (see the Supplemental Material). Thus, at the PbI$_2$-SnO$_2$ interface of LBSO/MAPbI$_3$, the perovskite forms of both materials are to be deformed into plumbic oxide. We find that this instability persists when La is doped on BSO (see the Supplemental Material). Apart from PbI$_2$-SnO$_2$ interface, the binding energies per unit cell of MAPbI$_3$/LBSO for all interfaces are strong $E_b$(PbI$_2$-BaO) = 4.38 eV/unit-cell, $E_b$(MAI-BaO) = 5.15 eV/unit-cell, and $E_b$(MAI-SnO$_2$) = 4.49 eV/unit-cell, larger than that of PbI$_2$- and MAI-terminated MAPbI$_3$/TiO$_2$ interface\cite{32}. The large binding energy indicates a strong hybridization among the interfacial atoms. Therefore, we expect that the interactions between two interfaces could significantly alter the individual electronic structures. 
	
	For MAI-terminations, the interaction between NH$_3$ of MA$^+$ and oxygen anion depends on the termination type of LBSO. At the SnO$_2$-termination, the short strong hydrogen bonding (SSHB)\cite{25,26} between NH$_3$ and O, where the hydrogen bounces back and forth between adjacent atoms, is present (see the Supplemental Material). This is rather an expected phenomenon as SSHB is also found at MAPbI$_3$/TiO$_2$ interface.\cite{27} The presence of such strong hydrogen bonding (HB) is crucial because it can further stabilize the interface and can affect the energy levels between the interfaces.\cite{27} At the BaO-termination, we observe a proton transfer from NH$_3$ to oxygen anion (see the Supplemental Material). We find that a slight La doping ($x = 3.7 \%$) on BSO hardly affects the optimized geometry (Fig. 2). The optimized geometries for the further doping ($x = 7.4, 11.0 \%$) are almost the same (see the Supplemental Material).
	
	Depending on the terminations of MAPbI$_3$ and BSO (PbI$_2$, MAI, SnO$_2$ and BaO), the work function of the system would vary significantly. This dependency on the interfacial morphology could then impact the practical performance of PSCs. However, the strong hybridizations between adjacent interfaces rearrange the pristine electronic density and so the expected large dependency on the terminations is removed (see the Supplemental Material). From the density of states (DOS) of LBSO/MAPbI$_3$ at PBE level, we observe a sizable hybridization of interfacial atoms irrespective of the termination types (see the Supplemental Material). We also note that these hybridizations are formed between ions from the conduction and valence states (i.e. Sn and I states) without impacting the relative conduction and valence levels of adjacent layers. Thus, the scenario such as the energy level shift by the induced dipole or the charge transfer will not be the dominant factor to describe the energy levels of LBSO/MAPbI$_3$ partly because of its role in compensating the two levels' difference and partly because of the sizable hybridization. Even though the calculation of the charge transfer is rather easy, we are not able to get the correct physics because the PBE energy levels are not accurate enough. An accurate calculation of the charge transfer is only possible using the PBE0+SOC+D3 calculation of the whole system, which is very demanding even for the supercomputers.
	
\begin{figure}
\centering
\includegraphics[width=8.6cm]{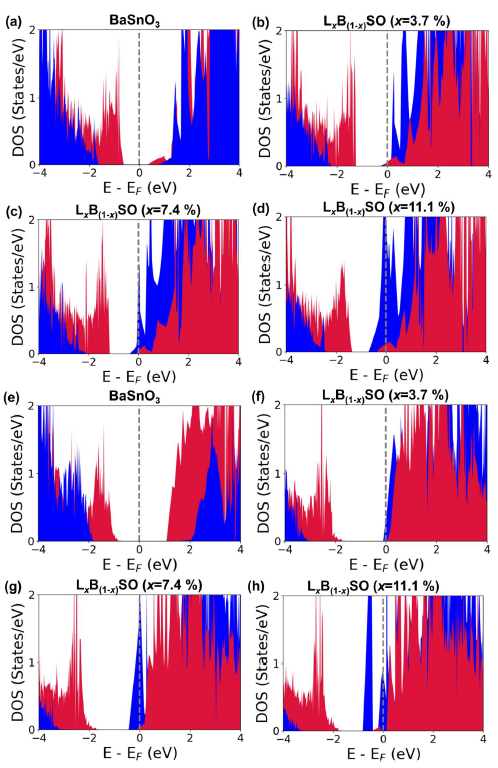}
\caption{\label{Figure. 3} Partial density of states (PDOS) of the MAI-SnO$_2$ terminated L$_x$B$_{(1-x)}$SO/MAPbI$_3$ ($x = 0, 3.7, 7.4, 11.1 \%$) interfaces with PBE0+SOC+D3 corrections, where (a) $x = 0 \%$, (b) $x = 3.7 \%$, (c) $x = 7.4 \%$, (d) $x = 11.1 \%$. PDOS of PbI2-BaO-terminated L$_x$B$_{(1-x)}$SO/MAPbI$_3$ with La doping of (e) $x = 0 \%$, (f) $x = 3.7 \%$, (g) $x = 7.4 \%$, and (h) $x = 11.1 \%$. The Fermi level $(E - E_F)$ is indicated as a gray dashed line. For clarity, we only plotted interfacial Pb $s, p$ (red) and Sn $s, p$ (blue) states, which are the main components of conduction and valence bands near EF. As the La doping $x$ increases, we notice that the electron energy barrier between MAPbI$_3$ and LBSO gradually decreases and is finally reversed. }
\end{figure}
	
	Given that the hybridization is strong in MAPbI$_3$/LBSO, it is crucial to check if this affects the electronic states of La and the conduction band minimum (Sn $s$, O $s$ states). For instance, if there exists a hybridization between two states, it would result in a significant modification on the conduction band's character and its energy levels. If this is true, then the assumed mechanism on the band gap renormalization would be incorrect. To confirm the mechanism in which the Coulomb interactions between La$^{3+}$ dopant and free electrons in the conduction band lowers the band edge energy, we checked the partial density of states (PDOS) of La in the slab (see the Supplemental Material). The mechanism in the bulk had been verified already in a previous study\cite{20}. We observe that regardless of the termination types, La $d$ states do not hybridize with Sn $s$ and O $s$ states at the interface. La $d$ states are 4 eV above and 2 eV above the CBM of LBSO at $x = 3.7 \%$, for SnO$_2$- and BaO-terminated cases, respectively (see the Supplemental Material). This is consistent at other La concentrations. Even for an extreme case where La dopant is at the very surface of LBSO slab, we observe no hybridization between La and Sn states. Therefore, we conclude that even though the hybridization between MAPbI$_3$ and LBSO exists, it does not affect La state, excluding a possibility of the hybridization with the conduction band. 
	
	To account for the energy levels at the LBSO/MAPbI$_3$ interface, we corrected PBE energy levels with PBE0 functional and its corrections are listed in the Supplemental Material. Without La doping, sizable electron barriers exist in all the three cases, 0.4 eV (MAI-SnO$_2$), 0.1 eV (MAI-BaO), and 0.45 eV (PbI$_2$-BaO), indicating that a pristine BSO would lose the open circuit voltage (Voc) and so it is not an ideal ETL for MAPbI3 (Fig 3(a) and 3(e)). To investigate the effect of La doping ($x = 3.7, 4.7, 11.1 \%$) on the LBSO/MAPbI$_3$ interface, we present the results of MAI-SnO$_2$ (Fig. 3a-c) and PbI$_2$-BaO (Fig. 3(d)-(f)) terminated interfaces. The MAI-BaO case can be found in the Supplemental Material. Upon La-doping $x = 3.7 \%$, only the conduction band level of LBSO shifts down (Fig. 3(b) and 3(f)) and the band alignment between MAPbI3 and LBSO starts to be favorable (or LBSO conduction band level becomes lower than MAPbI$_3$ conduction band level) for the electron transfer process. The energy shift is more pronounced for the further La-doping $x = 7.4$ and $11 \%$ (Fig. 3(c)-(d) and 3(g)-(h)). 
	
	For all the three interfaces, the band alignments of band edge states with respect to La-doping concentrations x ($\%$) are summarized in Fig. 4. For comparison, we have set the vacuum level of the system to 0 eV. In all the cases, an unfavorable conduction band alignment between MAPbI$_3$ and BSO can be overcome by a small amount of La doping. Therefore, as a function of La-doping $x$, one would find the optimal band alignment between MAPbI$_3$ and LBSO to maximize V$_{OC}$ and corresponding PCE of the device. Therefore, a further experimental work would be required to investigate the optimal La-doping range for the maximum photovoltaic efficiency in PSCs.
	
	In summary, we investigated the interface between a novel ETL material La-doped BaSnO$_3$ and the cubic MAPbI$_3$ PSCs using DFT with PBE0+SOC+D3. We found that the (001) LBSO/MAPbI$_3$ forms stable "all-perovskite" interfaces with large binding energy. Depending on the termination types, MA and O at the interface show SSHB and the proton transfer. Because LBSO can tune the conduction band energy selectively via La-doping due to the conduction band renormalization, it enables the control of band level solely by La-doping without halide or organic cation substitutions in PSC materials. This demonstrates an advantage of LBSO over other ETL materials in addition to the appreciated photostability over UV lights. We expect that the development of a novel ETL materials including LBSO would enable a further enhancement of the efficiency and the stability of the lead-halide-based PSC devices. 

\begin{figure}
\centering
\includegraphics[width=8.6cm]{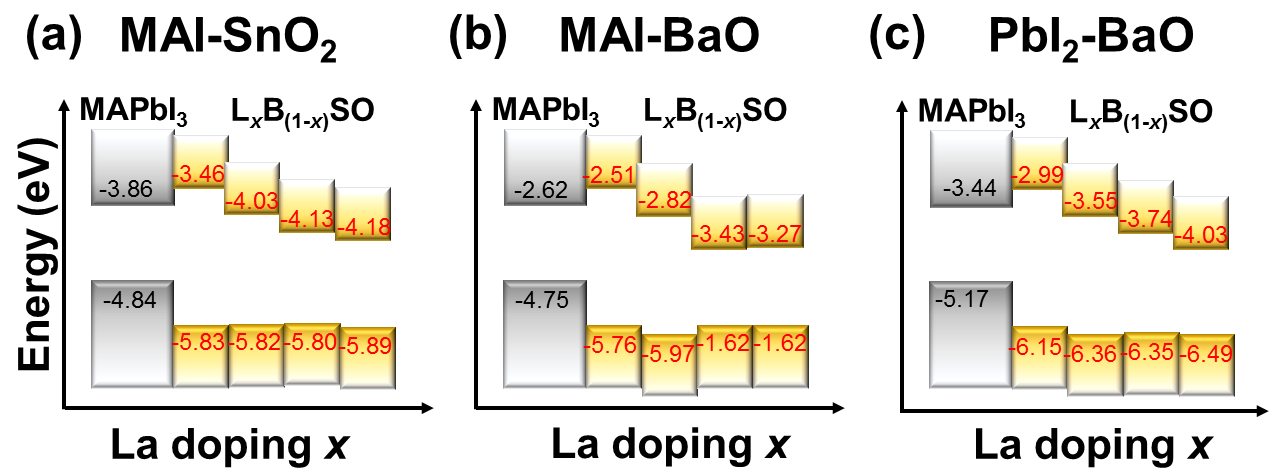}
\caption{\label{Figure. 4} The band alignment of (a) MAI-SnO$_2$, (b) MAI-BaO and (c) PbI$_2$-BaO terminated LBSO/MAPbI$_3$ interfaces at PBE0+SOC+D3 level of theory as a function of La doping x ($\%$). We have set the vacuum level to 0 eV. For all interface morphologies, the pristine BSO is not appropriate for ETL due to large electron barrier. La doping lowers the conduction level of BSO, thereby letting electrons transport to the electrode.}
\end{figure}

\begin{acknowledgments}
We thank Prof. S.-.I Seok and M.-J. Paik for their valuable discussions. C.W.M. conceived the idea, performed DFT calculation, and wrote the manuscript. All discussed the calculation results. K.S.K. revised the manuscript. This work was supported by National Honor Scientist Program (2010-0020414) and Basic Science Research Program (2015R1C1A1A01055922) of NRF. Computation was supported by KISTI (KSC-2017-S1-0025, KSC-2017-C3-0081).
\end{acknowledgments}

\bibliographystyle{apsrev4-1}
\bibliography{myungbib}

\end{document}